# Long-term investigations of radiocaesium activity concentrations in carps in north Croatia after the Chernobyl accident

**Zdenko FRANIĆ and Gordana MAROVIĆ**

Institute for Medical Research and Occupational Health, Ksaverska cesta 2

PO Box 291, HR-10000 Zagreb, Croatia

E-mail: franic@imi.hr

**Abstract**

Long-term investigations of radiocaesium activity concentrations in carps in the Republic of Croatia are presented. The radiocaesium levels in carps decreased exponentially and the effective ecological half-life of $^{137}$Cs in carps was estimated to be about 1 year for 1987-2002 period and 5 years for 1993-2005 period.

The observed $^{134}$Cs:$^{137}$Cs activity ratio in carps has been found to be similar to the ratio that has been observed in other environmental samples.

Concentration factor for carps (wet weight) was roughly estimated to be $128 \pm 74$ Lkg$^{-1}$, which is in reasonable agreement with model prediction based on K$^+$ concentrations in water.

Estimated annual effective doses received by $^{134}$Cs and $^{137}$Cs intake due to consumption of carps for an adult member of Croatian population are small, per caput dose for the 1987 – 2005 estimated to be $0.5 \pm 0.2$ μSv.

Due to minor freshwater fish consumption in Croatia and low radiocaesium activity concentrations in carps, it can be concluded that carps consumption was not a critical pathway for the transfer of radiocaesium from fallout to humans after the Chernobyl accident.

**Key words**:    carp; concentration factor; $^{137}$Cs; $^{134}$Cs; Chernobyl accident; dose

## 1. Introduction

Nuclear tests conducted in the atmosphere and releases of radioactive material from nuclear facilities are the main causes of man-made radioactive contamination of the environment. Once released to the atmosphere, long-range atmospheric transport processes can cause widespread distribution of such radioactive matter, although it may, as in the case of the Chernobyl accident, originate from a single point.

The resulting fallout, consisting of short and long-lived radionuclides, eventually impacts on humans, either directly or indirectly, by entering the food chain through plants and animals. In both cases radiation exposure represents a potential health hazard to the population through the direct irradiation and consumption of contaminated foodstuffs.

Among man-made radionuclides, those of caesium and strontium, particularly $^{137}$Cs and $^{90}$Sr, are regarded as a particular hazard to organisms. This results from the unique combinations of relatively long half-lives (30.14 and 29.12 years, respectively) and the chemical and metabolic properties of these radionuclides, which closely resemble potassium and calcium, respectively.

Investigations of the distribution and fate of natural, nuclear weapons-produced and reactor-released radionuclides in foodstuffs, including freshwater fish, have been conducted as a part of an extended and ongoing monitoring programme of radioactive contamination of the environment in Croatia (Popović, 1966-1978; Bauman et al., 1979 – 1992; Kovač et al., 1993 – 1998; Marović et al., 1999 - 2006). Regular measurements of radioactive contamination of freshwater fish started one year after the Chernobyl accident, in 1987.

Among the freshwater fish studied, special attention has been given to the common carp (*Cyprinus carpio*). As a widespread and rapidly breeding freshwater fish, carp species are fished and farmed across the world. In Croatia, carp represent the majority of the total freshwater fish catch and consumption. During the period 1991 – 2005, the total catch of carp in Croatia was about 55,000 tonnes or 3,800 ± 1,550 tonnes annually (Central Bureau of Statistics, 2000 – 2006). As virtually all this catch is consumed in Croatia, it implies an annual consumption of about 0.9 kg of carp per person. Unfortunately, data for carp catch and consumption for the 1987 – 1990 period are unavailable for the Republic of Croatia.

Common carp prefer large bodies of water with slow or standing water and soft, vegetated sediments. Since they can survive in stagnant and even polluted waters, which most other fish do not inhabit, carp represent good bioindicator species for freshwater radioecology.

## 2. Materials and methods

Carp samples were obtained once a year, usually in late spring or early summer, from commercial fish markets in the cities of Zagreb (45° 50' N, 16° 00' E) and Osijek (45° 30' N, 18° 40' E). Samples of river waters were also collected in the same cities, (the Sava river in Zagreb and the Drava river in Osijek), usually in late spring. A map showing the locations (cities of Zagreb and Osijek) of water and fish sampling sites is shown in Figure 1.

*FIGURE 1 ABOUT HERE*

From each site, 2 to 4 individual carp were obtained with masses of approximately 2 kg per individual. Therefore, they were all "large" (fresh weight more than 0.1 kg) according to the classification by Smith et al. (2002). The fish were cut into small pieces in order to obtain composite samples. Fish samples were dried in an oven and then ashed in a muffle furnace at 450 °C for 24 h. The $^{137}$Cs activity concentrations in carp are reported on a wet weight basis.

Samples of river waters were collected in the cities of Zagreb (the Sava river) and Osijek (the Drava river), usually in late spring. Total $^{137}$Cs activity concentrations, i.e. dissolved and solid forms, were determined in water samples. Prior to analysis, samples of 50 L were evaporated to 1 L volume in order to concentrate the $^{137}$Cs content. It was not possible to collect water samples from the local fishponds in which carp were farmed and it is assumed that water from adjacent rivers is a good representation of surface waters in the study areas.

Fallout samples were collected monthly in the city of Zagreb at the Institute for Medical Research and Occupational Health (45° 50' 7.3" N, 15° 58' 58.7" E). Funnels, which were used for rainwater collection, had a 1 m$^2$ collection area. Precipitation was measured by a Hellman pluviometer. As in the case of river water samples, rainwater was evaporated to a volume of 1 L in order to concentrate the $^{137}$Cs activity concentration prior to analysis.

Gamma-ray spectrometry systems based on a low-level ORTEC Ge(Li) detector (FWHM 1.87 keV at 1.33 MeV $^{60}$Co and relative efficiency of 15.4% at 1.33 MeV) and ORTEC HPGe detector (FWHM 1.75 keV at 1.33 MeV $^{60}$Co and relative efficiency of 21% at 1.33 MeV) coupled to a computerized data acquisition system were used to determine radiocaesium and $^{40}$K levels in the samples from their gamma-ray spectra. Ash from the fish samples was packed in cylindrical plastic containers of appropriate volume, which were placed directly on the detector. River water and fallout samples were measured in Marinelli beakers. Counting times depended on sample activities, ranging from 10,000 to 250,000 seconds, typically being 80,000 s.

Quality assurance and intercalibration measurements were performed through participation in International Atomic Energy Agency (IAEA) and World Health Organization (WHO) international intercalibration programmes, which include regular checks on blank and background samples as well as quality control measurements.

Radiocaesium activity concentrations in fish samples in this paper were reported as averages of two sampling locations, which assumes that the water bodies in both locations had similar characteristics. The relative error between both locations never exceeded 25 %.

**3. Results and discussion**

*3.1. $^{137}$Cs activity concentrations in carp*

Fallout from highly radioactive atmospheric plumes originating from the damaged Chernobyl nuclear reactor was spread and transported all over Europe. Fortunately, due to the prevailing meteorological conditions at the time after the accident, Croatia was only partially affected by the edge of one of the plumes (United Nations Scientific Committee on the Effects of Atomic Radiation, 1988), as indicated in Figure 2.

*FIGURE 2 ABOUT HERE*

The highest observed $^{137}$Cs activity concentrations in Chernobyl fallout were recorded in May 1986, resulting in a surface deposition of 6200 Bq m$^{-2}$ (Bauman et al., 1978 – 1991;

Franić, 1992a). The highest $^{137}$Cs activity concentration recorded in carps was 19.5 ± 1.5 Bq kg$^{-1}$ in 1987, while in 2004 the value was only 0.07 ± 0.01 Bq kg$^{-1}$. The latter value is comparable with the $^{137}$Cs activity concentration found in the late 1990s for trout (Bauman et al., 1978 - 1991) although no data for $^{137}$Cs activity concentrations in trout for the 1980s and early 1990s are available.

*3.2. Effective ecological half-life of $^{137}$Cs in carp*

To study the time course the $^{137}$Cs activity concentrations and to assess the effective ecological half-life of $^{137}$Cs in carp, measurement data have been graphed as a function of time. The distribution shows a relatively rapid decrease during the first six years after the Chernobyl accident. After approximately six years, this decrease was significantly less rapid. The same pattern has been observed for other fish species (Smith et al., 2000a; Hanslik et. al., 2005) as well as various other foodstuffs (Schwaiger et al., 2004).

A first order kinetic equation was used to parameterise time changes in the $^{137}$Cs activity concentrations in carp. Therefore, for the periods of 1987-1992 and 1993-2005 the measured data were fitted to the following exponential function:

$$A_{carp}(t) = A_{carp}(0)\, e^{-kt} \qquad (1)$$

where:

$A_{carp}(t)$ is time-dependent activity concentration of $^{137}$Cs in carp (Bqkg$^{-1}$ wet weight),

$A_{carp}(0)$ initial activity concentration of $^{137}$Cs in carp (Bqkg$^{-1}$ wet weight) and

$ln(2)/k = T_{1/2,eff}$ effective (observed) ecological half-life of $^{137}$Cs in carp (years).

In the immediate period after the Chernobyl accident, i.e. 1987 –1992 the observed effective ecological half-life of $^{137}$Cs in carp was about 1.05 years (r = 0.92, P(t) < 0.01 for 4 degrees of freedom). However, the effective ecological half-life of $^{137}$Cs in carp increased to 5.05 years for the 1993 –2005 period (r = 0.82 with P(t) < 0.001 for 11 degrees of freedom). The observed and modeled $^{137}$Cs activity concentrations in carp are shown in Figure 3.

FIGURE 3 ABOUT HERE

In order to obtain the standard deviation of $T_{1/2,eff}$, Monte Carlo simulations were performed. To ensure conservative estimates, as well as to simplify calculations, a uniform distribution has been assumed over the A ± 2σ value of $^{137}$Cs activity concentrations in carp for respective years, although a normal distribution would perhaps be more realistic. For each year a random value was generated over the interval [A - 2σ, A + 2σ] and a ln(2)/k value estimated by fitting equation (1) to the stochastically generated data. This process was repeated 100 times and 100 values for ln(2)/k = $T_{1/2,eff}$ were thus obtained. From this data set the mean values and standard deviations for $T_{1/2,eff}$ were calculated to be 1.05 ± 0.06 years for 1987 – 1992 period and 5.05 ± 0.32 years for 1993 – 2005 period.

These results are in good agreement with literature data. Hanslik et al. (2005) reported the observed $^{137}$Cs half-lives for freshwater fish species (both piscivorous and herbivorous) in the Vltava River basin (Bohemia) to be about 1 year during 1986 – 1992 and 5.1 years during 1994 – 2002. The effective ecological half-life for freshwater fish in the post Chernobyl period in Nordic countries was discussed in the Nordic nuclear safety research report NKS-123 (Nielsen and Andersson, 2006). During 1988-2002 the observed ecological half-life of $^{137}$Cs in perch in Finnish lakes varied from 3 to 9 years while half-lives in trout and in Arctic char varied from 1.4 to 4.7 years during 1988 – 2004. Trout in Norwegian lakes experienced a more rapid reduction in $^{137}$Cs burdens compared with Finnish lakes, the highest observed ecological half-life, determined for the time period 1989 – 2000, being 6.3 years. Smith et al. (2000a) reported effective biological half-life of $^{137}$C in young fish, water and terrestrial vegetation during the first five years after the Chernobyl accident to be 1 – 4 years. However, after that period, the effective biological half-life of $^{137}$C increased to 6 – 30 years.

The effective ecological half life results from biological elimination of $^{137}$Cs from fish as well as radioactive decay. In order to estimate the 'true' ecological half-life the effective constants k should be corrected for radioactive decay. Therefore constant k from the equation (1) can be written as:

$$k = \lambda + k_R \tag{2}$$

where $\ln(2)/\lambda = 30.14$ y is the physical half-life of $^{137}$Cs and $\ln(2)/k_R = T_{1/2,e}$ is ecological half-life for $^{137}$Cs.

From equation (2) the 'true' ecological half-lives for $^{137}$Cs in carp $T_{1/2,e}$ were found to be $1.09 \pm 0.06$ and $6.06 \pm 0.47$ years, which are slightly higher than the effective ecological half-lives for the respective 'fast' and 'slow' periods.

## 3.3. $^{137}$Cs : $^{134}$Cs activity ratio in carp

The presence of $^{134}$Cs (half-life of 2.06 years) in the Croatian environment was detected for the first time in May 1986. The estimated amount of caesium released after the reactor explosion at Chernobyl was $3.7 \times 10^{16}$ Bq of $^{137}$Cs (13% of total reactor inventory) and $1.9 \times 10^{16}$ Bq of $^{134}$Cs (10% of total reactor inventory (International Atomic Energy Agency, 1986). Thus, the initial value for the $^{134}$Cs:$^{137}$Cs activity ratio in May 1986 was 0.51. This activity ratio was not altered during the passage of the radioactive plume from Chernobyl to other geographical areas in Europe.

As the half-life of $^{137}$Cs (30.14 years) is about 15 times longer than that of $^{134}$Cs, the $^{134}$Cs:$^{137}$Cs activity ratio R(t) should decrease, in a predictable way, due to differential radioactive decay, according to the following relationship:

$$R(t) = \frac{1.9 \times 10^{16}}{3.7 \times 10^{16}} \times e^{\ln(2) \times t \times (\frac{1}{T_1} - \frac{1}{T_2})} \tag{3}$$

where:

t is time elapsed after the Chernobyl accident and

$T_1$ and $T_2$ are the physical half-lives for $^{137}$Cs and $^{134}$Cs, respectively.

In 1987 the observed $^{134}$Cs:$^{137}$Cs activity ratio in carp was $0.41 \pm 0.06$, compared with the theoretical prediction of 0.38. In 1988 that ratio was $0.27 \pm 0.04$ (Figure 4).

*FIGURE 4 ABOUT HERE*

Similar ratios, decreasing according to equation (3), have been found in most other environmental samples (Bauman et al., 1987 - 1992; Kovač et al., 1993 - 1998; Marović et al., 1999 – 2004; Franić et al. 2006) the only exception being mushrooms (Franić et. al, 1992b). In mushrooms, excess $^{137}$Cs from pre-Chernobyl fallout, affected the $^{134}$Cs:$^{137}$Cs concentration ratios. As $^{134}$Cs decayed and migrated to deeper layers within soils, the observed $^{134}$Cs:$^{137}$Cs concentration ratios approached the values theoretically predicted.

In 1990 and afterwards, contamination of carp by $^{134}$Cs originating from the Chernobyl accident was detectable only at a very low level.

*3.4 Concentration factors*

Radionuclides introduced into natural waters undergo various biological processes and may enter complicated food-webs. Therefore, a variety of mathematical models have been developed to predict the behavior and fate of radionuclides in fresh water ecosystems, as described by IAEA (2000) and Monte et al. (2003).

Since one of the routes of human exposure to artificial radionuclides from the freshwater environment is through ingestion of contaminated organisms, it is interesting to investigate the bioaccumulation of $^{137}$Cs in carp. The level of radioactive contamination of aquatic biota by specific radionuclides is usually defined in terms of a concentration factor (CF) which is defined as the ratio of activity concentration in a biological material to the activity concentration in the ambient water:

$$CF_m = \frac{A_m(organism)}{A_m(water)} \qquad (4)$$

where:

*CF$_m$*     is the concentration factor of radionuclide *m* for a specific organism (L kg$^{-1}$)

*A$_m$ (organism)*     is the activity concentration of radionuclide *m* in that organism (Bq kg$^{-1}$ wet weight) and

*A$_m$ (water)*     s the activity concentration of radionuclide *m* in water (Bq L$^{-1}$).

Use of concentration factors assumes equilibrium between organisms and water, which may not be the case in the real environment due to complex inputs of radionuclides and the variability in parameters influencing activity concentrations in abiotic compartments. Concentration factors do not have any mechanistic basis, simply representing the ratios of radionuclide activity concentrations between biological material and ambient water. Nevertheless, if the concentration factor of a specific radionuclide is known for a given organism, the level of the radionuclide activity concentration in this organism can be predicted, based on its activity concentration in the ambient water.

Since carp were obtained in fish markets in the cities of Zagreb and Osijek, without knowing the exact locations of the fish farms from which they came, CRs were calculated using $^{137}$Cs activity concentrations observed in the rivers Sava and Drava. The calculated value of CF for $^{137}$Cs in North Croatian carp is $128 \pm 73$ L kg$^{-1}$. If the data for Zagreb and Osijek are analyzed separately, respective CFs are $129 \pm 64$ and $125 \pm 82$ L kg$^{-1}$. The small deviations in these CF estimates is expected given the assumptions made. It should be noted that non-predatory fish, like carp, are expected to have lower CFs for radiocaesium than predatory fish (Kryshev, 1995). Nevertheless, these CFs correspond with other reported CF values for $^{134}$Cs and $^{137}$Cs, which fall in the range $3 \times 10^1 - 3 \times 10^3$ (IAEA, 1994). Similar ranges of CFs have also been observed by other authors. Strand et al. (1999) reported CFs for selected fresh water fish (pike, perch and roach) in artificial reservoirs in the southern Urals, (Russia) in the range of 240 –1400 L kg$^{-1}$. Smith et al. (2000b, 2002) have reported fish-water CFs for perch, rudd, roach, gudgeon, ruffe, bream and goldfish in the range $10^2 - 10^4$ L kg$^{-1}$ in 10 lakes in Russia, Belarus and Ukraine. Hanslik et al. (2005) have reported somewhat greater fish-water CFs for the fish sampled in Orlik Reservoir (the Vltava river basin), in the range of 1000 – 2000 L kg$^{-1}$.

Another important factor which affects rates of uptake and elimination of radiocaesium in fish, and therefore CFs, is the influence of potassium as well other chemical parameters (IAEA, 1994; Poon and Au, 1999; Smith et. al., 2002; Smith et al., 2005). Specifically, fish-water CFs are inversely proportional to potassium concentrations and expressions have been derived to describe this relationship (Coughtrey and Thorne, 1983; IAEA, 1994; Smith et al., 2005).

For non-predatory fish, like carp, it has been observed (Smith et al., 2005) that:

$$CF \approx 61.3 / [K^+] \tag{5}$$

where:

*CF*  is the fish-water CF for $^{137}$Cs in m$^3$ kg$^{-1}$ and

*[K$^+$]*  is the potassium concentration measured in μmol L$^{-1}$.

The potassium concentrations in the waters of the Drava and Sava rivers, estimated from gamma spectrometric data for $^{40}$K, were in the range 170 - 250 μmol L$^{-1}$. Similar potassium concentrations, 6.7 mg L$^{-1}$ (≈ 170 μmol L$^{-1}$) have been observed elsewhere (Neal and Robson, 2000). Therefore, using equation (5), CFs would fit into the range 245 – 370 L kg$^{-1}$ which is slightly higher than the observed CF values for carp. It should be noted, however, that potassium concentrations in some Croatian lakes, as well as ponds, is significantly higher than in the Sava and Drava rivers (Marović et al., 1999 – 2006) while $^{137}$Cs activity concentrations are quite similar. Consequently, the related CF values for carps are likely to be lower than the ones predicted by equation (5) using the data from the Sava and Drava rivers.

3.5  *Dosimetry*

Due to the relatively high contribution of ingestion dose to total dose received by the Croatian population after nuclear fallout (Lokobauer et al., 1998), a reliable knowledge of ingestion dose is of particular importance. Data on activity concentrations of $^{134}$Cs and $^{137}$Cs in carp allow us to estimate the doses incurred by consumption of this fish species. Dose conversion factors, i.e. effective dose per unit intake via ingestion for adult members of the public, are $1.3\times10^{-8}$ SvBq$^{-1}$ and $1.9\times10^{-8}$ SvBq$^{-1}$ for $^{137}$Cs and $^{134}$Cs respectively (IAEA, 1996). As the ratio of dose conversion factors for $^{137}$Cs and $^{134}$Cs is ≈ 0.7, it implies that ingestion of $^{134}$Cs contributes about 30% more to the dose, compared with ingestion of the same activity concentration of $^{137}$Cs.

In the Republic of Croatia, overall consumption of fish, including fresh water and sea species, is relatively small, being between 7 and 8 kg per year per person (Central Bureau of Statistics, 2000 – 2006). The collective effective dose incurred due to food consumption over a specific time period depends on the activity concentration within the food and on the quantity of that food which is consumed. The dose can be expressed as:

$$E = C \sum_m D_m^{cf} A_m \qquad (6)$$

where:

$E$    is the effective dose in Sv,

$C$    is the total annual *per capita* consumption of food,

$D_m^{cf}$    is the dose conversion factor for radionuclide $m$, i.e. effective dose per unit intake, which converts the ingested activity to effective dose and

$A_m$    is the mean annual specific activity of radionuclide $m$ in food (Bqkg$^{-1}$).

Based on the statistical data for the consumption of carp during 1991 – 2005 (3800 ± 1550 tonnes annually) the same consumption rate was assumed for the period 1987 – 1990 as well. The estimated collective effective dose due to $^{137}$Cs and $^{134}$Cs ingestion by carp consumption for the Croatian population (4.5 ×10$^6$ inhabitants) during 1987-2005 is 2.2 ± 0.9 person-Sv. 1.5 ± 0.6 person-Sv can be attributed to $^{137}$Cs and 0.7 ± 0.3 person-Sv to $^{134}$Cs. The majority of the overall dose was incurred during the 1987 (1.0 ± 0.4 person-Sv and 0.6 ± 0.2 person-Sv for $^{137}$Cs and $^{134}$Cs, respectively).

The annual effective collective doses are shown in Figure 5.

*FIGURE 5 ABOUT HERE*

It should be noted that the doses were not corrected to take into account inedible parts of the fish such as bones and scales. Nevertheless, these are rather small doses, since for the 1987 – 2005 period per capita dose is . 0.5 ± 0.2 µSv. However, the doses are small not only because of low activity concentrations of radiocaesium in carp, but also due to the low consumption of freshwater fish in Croatia. In summary, it can be argued that freshwater fish consumption is not a critical pathway for human intake of radiocaesium from the environment.

## 4. Conclusions

A relatively rapid decrease in $^{137}$Cs activity concentrations in carp was observed for the immediate post-Chernobyl period (1987 – 1992), the effective ecological half-life of $^{137}$Cs in carp within this period being approximately 1 year. However, approximately 6 years after the accident, this rate of decrease slowed significantly and the effective ecological half-life of $^{137}$Cs in carp increased to 5 years, the same pattern being observed for some other fish and other foodstuffs.

The concentration factor (CF) for carp (wet weight) was estimated to be $128 \pm 74$ L kg$^{-1}$, a value which is in reasonable agreement with model predictions based on K$^+$ concentrations in water as well as with CF values observed elsewhere.

The activity concentrations of $^{137}$Cs in carp became quite low within a few years after the Chernobyl nuclear accident, while $^{134}$Cs activity concentrations were generally below detection limit after 1990. Consequently, only very low doses have been received by the general population as a result of freshwater fish consumption and it can be concluded that carp consumption was not a critical pathway for the transfer of radiocaesium to humans after the Chernobyl accident.

**Acknowledgements**

This study is a part of a research project *Radioecology of the Adriatic Sea and Coastal Areas*, supported by the Ministry of Science, Education and Sports of the Republic of Croatia.

The authors wish to thank Mrs. Ljerka Petroci for her excellent technical assistance and help.

**Figures**

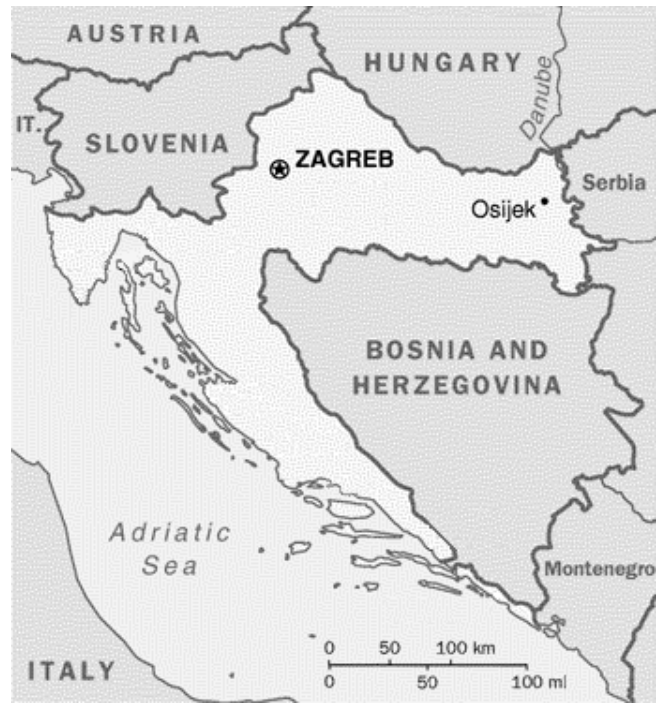

**Figure 1**

Map showing the water and fish sampling sites (cities of Zagreb and Osijek).

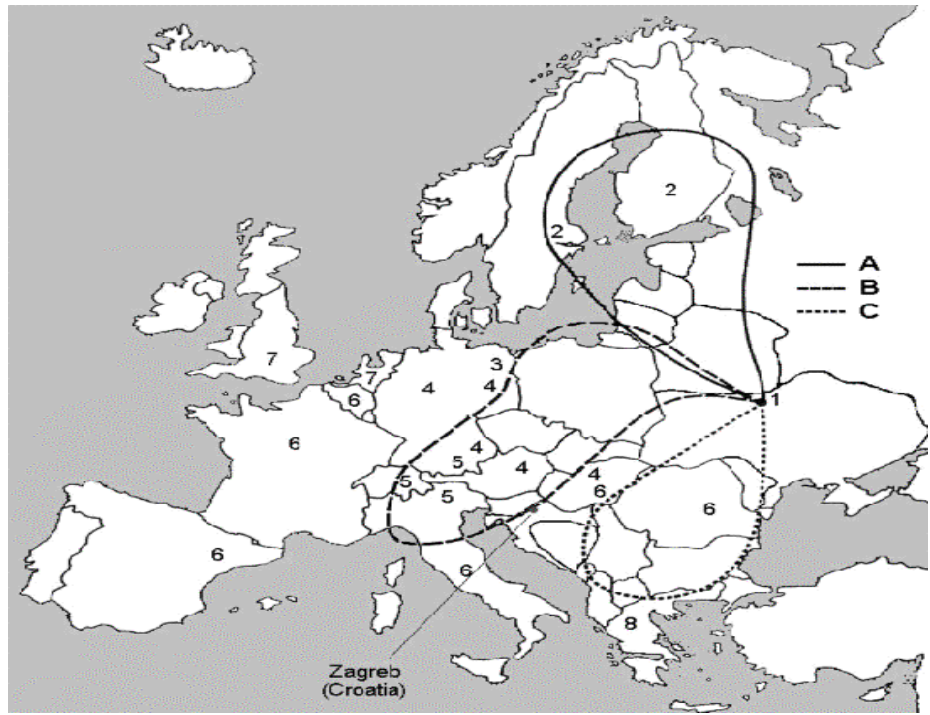

**Figure 2**

Spreading of radioactive plumes over the Europe after the Chernobyl nuclear accident. Numbers 1-8 represent plume arrival times at respective areas: 1 = April 26, 2 = April 27, 3 = April 28, 4 = April 29, 5 = April 30, 6 = May 1, 7 = May 2 and 8 = May 3. The figure has been adapted from the UNSCEAR report for 1988.

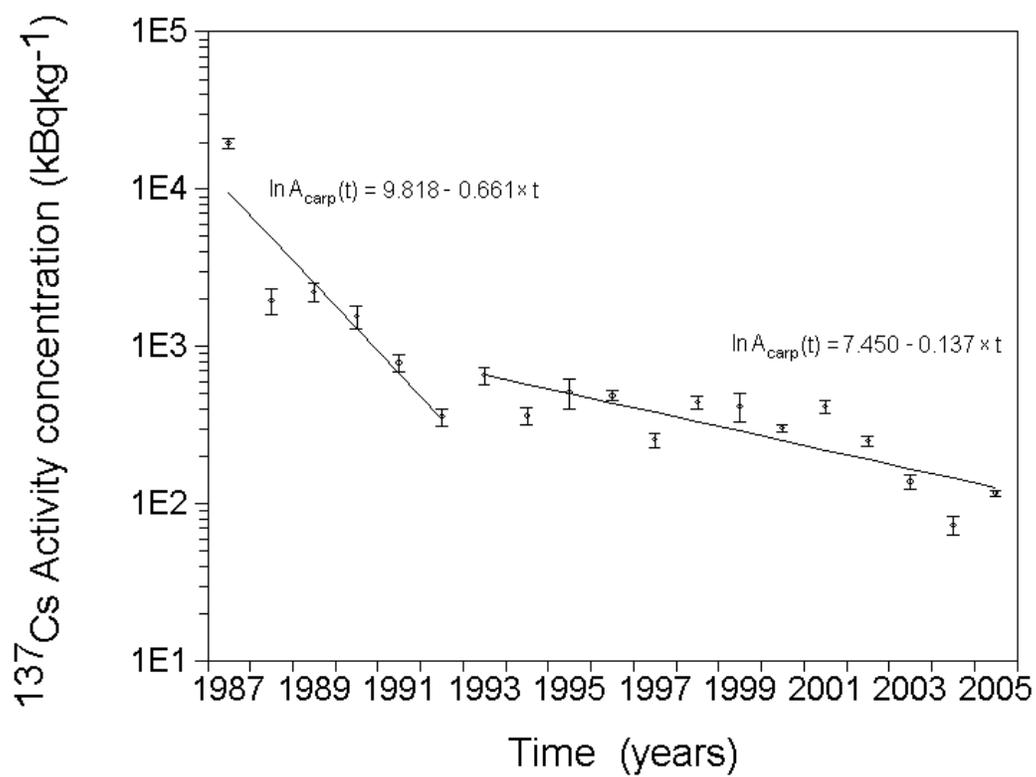

**Figure 3**

Observed and modeled $^{137}$Cs activity concentrations in carp as a function of time during 1987 −1992 and 1993 – 2005.

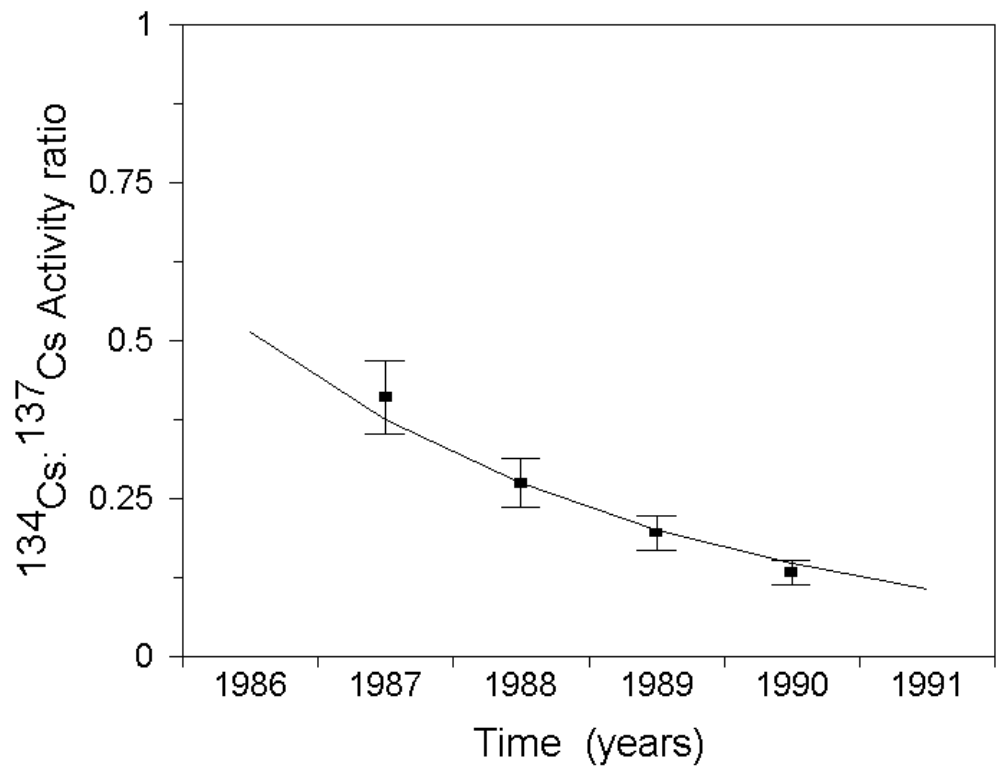

**Figure 4**

Observed and predicted $^{134}$Cs:$^{137}$Cs activity concentration ratio in carp.

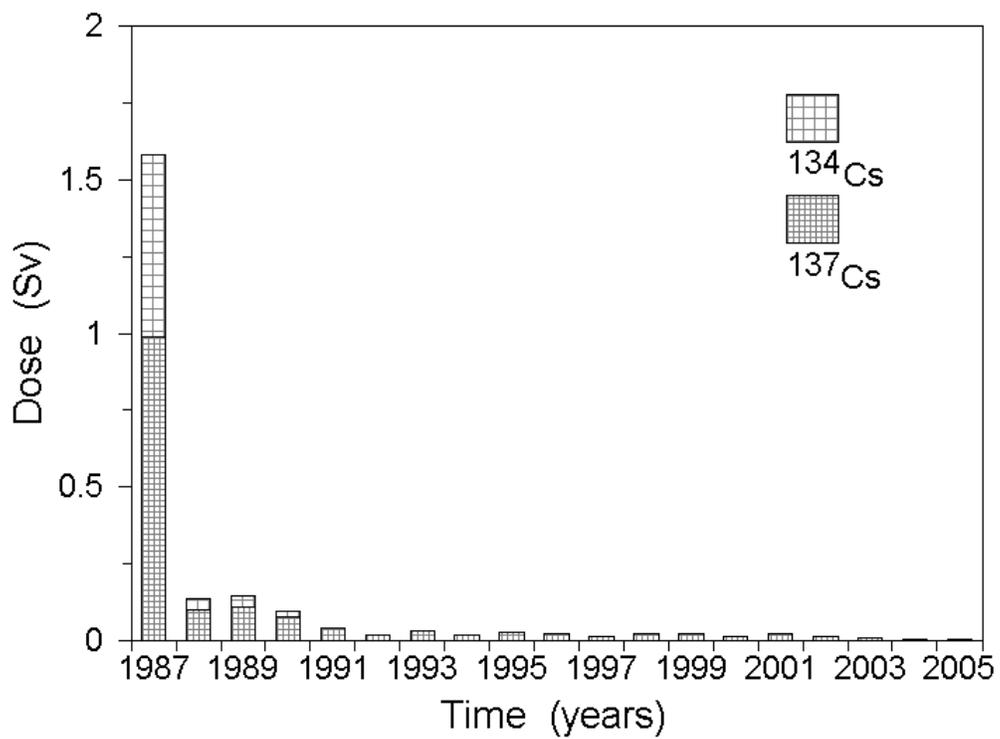

**Figure 5**

Estimated annual collective effective doses due to $^{134}$Cs and $^{137}$Cs intake due to consumption of carp by the Croatian population.